\documentclass[dvips]{article}
\usepackage{graphicx}
\usepackage{amssymb}

\def\<{\langle}
\def\>{\rangle}

\def\bv{{\bf b}}

\def\qv{{\bf q}}
\def\rv{{\bf r}}

\def\vv{{\bf v}}

\def\Bv{{\bf B}}
\def\Ev{{\bf E}}
\def\Fv{{\bf F}}
\def\Gv{{\bf G}}

\def\Xv{{\bf X}}
\def\Vv{{\bf v}}
\def\zu{\hat{\bf z}}

\def\bt{{\bf B}_T}
\def\et{{\bf E}_T}
\def\bl{B_z}
\def\el{E_z}

\def\K{{\cal K}}

\def\lambdabar{\lambda\raise0.4ex\hbox{\kern-0.5em\hbox{--}}\ }

\def\be{\begin{equation}}
\def\ee{\end{equation}}

\begin{document}
\Large
\centerline{INTERFERENCE AND SHADOW EFFECTS IN THE PRODUCTION}
\centerline{OF LIGHT BY CHARGED PARTICLES IN OPTICAL FIBERS%
\footnote{Presented at the International Conference{\it Radiation by Relativistic Electrons in Periodic Structures} (RREPS'O7, Prague, Sept. 24-27, 2007.}
}
\normalsize
\vskip 0.6 true cm

\centerline{Xavier ARTRU\footnote{e-mail: x.artru@ipnl.in2p3.fr}
and C\'edric Ray 
}

\centerline{\it Institut de Physique Nucl\'eaire de Lyon, Universit\'e de Lyon,}

\centerline{\it CNRS- IN2P3 and Universit\'e Lyon 1, 69622 Villeurbanne, France}

\vskip 0.8 true cm
\centerline{\bf ABSTRACT}

\medskip\noindent
A charged particle passing through or near a narrow optical fiber induces, by polarisation, coherent light guided by the fiber. In the limit of zero crossing angle, the radiation tends towards a Cherenkov radiation with a discrete spectrum, studied by different authors. 
If the particle crosses a bent fiber at regularly spaced points, interference gives quasi-monochromatic lines. 
If the particle passes near an end of the fiber, light is produced by the capture of virtual photons through the end face. 
An alternative way consists in sticking a metallic ball to the fiber: the passing particle induces plasmons which are then evacuated as light in the fiber. Interferences can occur between lights from several ends or balls. Applications of these various light signals to beam diagnostics are discussed. 
The shadow effect, which reduces the photon yield when the particle runs parallel to a row of balls, is pointed out and an upper bound $-dE/dz\le C(Ze/b)^2$ for the particle energy loss is conjectured ($Ze$ is the particle charge, $b$ the impact parameter and $C$ a numerical constant).
This bound should also apply to other kinds of light sources, in particular to Smith-Purcell radiation.

\medskip\noindent
{\bf keywords:}
particle detector, beam diagnostics, optical fibers, Cherenkov radiation, Smith-Purcell, plasmon

%41.85.Qg=particle beam analysers and monitors, 29.40.-n=radiation detectors, 41.60.Bq=Cherenkov radiation, 42.55.Wd=other optical fiber devices

\section{Introduction}\setcounter{equation}{0}

In a previous paper \cite{Frascati06}, we have studied the production of coherent light inside a thin optical fiber by a charged particle passing through or near the fiber. This light does not result from scintillation of the medium but from its transient polarisation under the field of the particle. In that respect, it belongs to the same family (\emph{polarisation radiation}) as Cherenkov light and transition radiation and, indeed, for a broad enough fiber \cite{DIRC,Gorodetsky} one can treat it as Cherenkov radiation. Taking into account the finite path in the medium it is more precisely described as transition radiation and has no velocity threshold. However in the case of a narrow fiber the curvature of the surface is too strong and one cannot use the transition radiation formula. Besides, radiation is produced even when the particle does not penetrate inside the fiber. This phenomenon is not taken into account by transition radiation but occurs, for instance, in the Cherenkov radiation ``at a distance'' produced by a particle moving parallel to the boundary of a dielectric medium. 

A solution $\{\Ev(\rv,z,t),\Bv(\rv,z,t)\}$ of the Maxwell equations in the presence of the fiber can be decomposed into frequencies $\omega$ and an infinite number of transverse \emph{modes}, labeled by  $m=\{M,\nu,{\rm sign}(K)\}$~: 
\be\label{decompose-mode}
\pmatrix{\Ev(\rv,z,t)\cr\Bv(\rv,z,t)}= \int^{\infty}_{-\infty} \frac{d\omega}{2\pi}
\sum_{m} c{(\omega,m)} \ e^{iK(\omega,m)z-i\omega t}  \  \pmatrix{\Ev_{\omega,m}(\rv)\cr\Bv_{\omega,m}(\rv) } \,.
\ee
$\rv=(x,y)$ is the transverse coordinate, $K=k_z$ the longitudinal momentum, $M=J_z$ the angular momentum  and $\nu$ the radial quantum number. Omitting the sign of $K$, we will write $m=\{M,\nu\}$. $\,K$ and $\omega$ are linked by the dispersion relation
\begin{equation}
K=K(\omega,m)\,.
\label{disp}
\end{equation} 
Strictly speaking, the radial quantum number $\nu$ takes discrete values for guided modes and continuous values for scattering modes. However, quantising the field in a large cylinder, $\nu$ is treated in (\ref{decompose-mode}) as a discrete index. If the fiber is sufficiently narrow, there is only one or a few number of guided modes. At very small thickness, only the lowest mode (lowest $\omega$ at fixed $K$ or highest $K$ at fixed $\omega$), characterised by $M=\pm1$, $\nu=1$ and called $HE_{11}$, survives.  

The modes can be excited, by polarisation, during the passage of a charged particle through or near the fiber.
Contrary to the scintillation light, this \emph{polarisation radiation} is coherent: the emitted modes have definite and calculable phases and one can make interferences between lights produced at several crossing points between the fiber and the particle. If there are many crossings with a spatial periodicity, for instance with a periodically bent fiber, the interference builds up narrow peaks in the spectrum. 

The case of straight and parallel trajectory and fiber has been treated particularly  by Bogdankevich and Bolotovskii \cite {Bog-Bol}, Zhevago and Glebov \cite {Zhev-Gleb-Cherenkov}. A guided \emph{fiber Cherenkov} radiation is produced, with a discrete spectrum given by the condition that the velocity $v_z$ of the particle is equal to the phase velocity $v_{ph}$ of the mode:
\begin{equation}
v_z = v_{ph}(\omega,m)\equiv \omega/K(\omega,m) \,.
\label{Cher}
\end{equation} 
The radiated energy is proportional to the common straight section $L$ between the particle trajectory and the fiber. At nonzero but small angle $\theta$ between the trajectory and the fiber, it is proportional to $L_{eff}\sim a/\theta$, where $a$ is the fiber radius.

In the fiber Cherenkov mechanism, or the one studied in \cite{Frascati06}, the radiation is produced in a continuous part of the fiber. 
Polarisation radiation can also be produced by a charged particle passing close to one end of the fiber (\emph{cut fiber scheme}), or close to a bump or indentation of its surface. A particularly interesting kind of bump is a metallic ball sticked to the surface (\emph{ball scheme}). In this case a plasmon may be produced in the ball \cite{Zhev-plasmon,Garcia+Howie,Garcia+Y+A} and evacuated as guided light in the fiber. 
Interference effects can also be obtained in these two cases: light produced in several cut fibers can be gathered in a single fiber through junctions and interfere there; light produced at periodically spaced metallic ball sticked to one fiber can interfere making narrow peaks in the spectrum. 

In this paper we will discuss the interference effects which could be used for a particle detector or for beam diagnostics. In section 2 we review the main results of \cite{Frascati06}, give the explicit formula for the $\theta\to0$ limit and recall the formula giving the interference peaks when the particle trajectory and the fiber cross each other periodically. 
In section 3 we consider the production of polarisation radiation at one end of the fiber or via a plasmon in a metallic ball and the possiblity of interferences. A possible reduction of the intensity by a ``shadow effect'' is outlined in section 4 and an upper bound is conjectured for the total energy loss per unit length.

\section{Production of light in a continuous part}\setcounter{equation}{0}
We consider a narrow cylindrical fiber parallel the $z$-axis. A particle of charge $Ze$ is {\it crossing} or {\it passing near} the fiber along the straight or curved trajectoty $\Xv(t)$. The amplitude $R$ of spontaneous emission in the mode $(\omega,m)$ is given by%
\footnote{We use relativistic quantum units where $\hbar=c=1$ and rational Maxwell equations, like $\nabla.\Ev=\rho$ without a $4\pi$ factor. $\ e^2/(4\pi)=\alpha=1/137$.}
\begin{equation}
R_{\omega,m} = {iZe\over\omega} \int d\Xv(t)\cdot \Ev^*_{\omega,m} (\rv) \ e^{i\omega t-iK(\omega,m)z} 
\label{ampli}
\end{equation}
and the photon distribution reads
\begin{equation}
d{\cal N}_{\omega,m} = \frac{\omega \ d\omega}{2\pi P} \ \left| R_{\omega,m} \right|^2\,,
\label{distrib'}
\end{equation}
where $P$ is the \emph{power} of the mode, that is to say the flux of the Poynting vector through a plane perpendicular to the fiber:
\be\label{flux}
P = \int d^2\rv \ \Re \left\{\Ev^* \times \Bv\right\}_z.
\ee
The expression (\ref{distrib'}) is invariant under a change of the normalisation of the mode fields. 

We assume for simplicity that the fiber has uniform refraction index $\sqrt{\varepsilon}$ and no cladding.
Inside the fiber, the photon has a real transverse momentum $q$. Outside the fiber, the guided mode is an evanescent wave $\sim e^{-\kappa r}$ corresponding to an imaginary transverse momentum $i\kappa$.  $q$ and $\kappa$ are related to $\omega$ and $K$ by
\be
q(\omega,m) = \left\{ \varepsilon\omega^2 - K^2 \right\}^{1/2}\,,
\qquad
\kappa(\omega,m) = \left\{ K^2 - \omega^2 \right\}^{1/2}.
\ee
Due to the evanescent wave, $P$ contains an external part, the relative size of which is plotted in Fig.1 versus the dimensionless frequency parameter $\omega.a$, for the lowest mode $HE_{11}$ in a fused silica fiber ($\sqrt{\varepsilon}=1.41$). In this figure is also plotted the phase velocity $v_{ph} = \omega/K$, which is intermediate between the velocity in the bulk medium, $c_{med}=1/\sqrt{\varepsilon}$, and that in vacuum. 
The explicit form of the dispersion relation, the derivation of (\ref{ampli}-\ref{distrib'}) and some numerical results are given in \cite{Frascati06}, for an uniform index and a straight particle trajectory. 
\begin{figure}[htbp] 
	\centering
\includegraphics*[scale=1.,clip,bb=10 10 330 230]{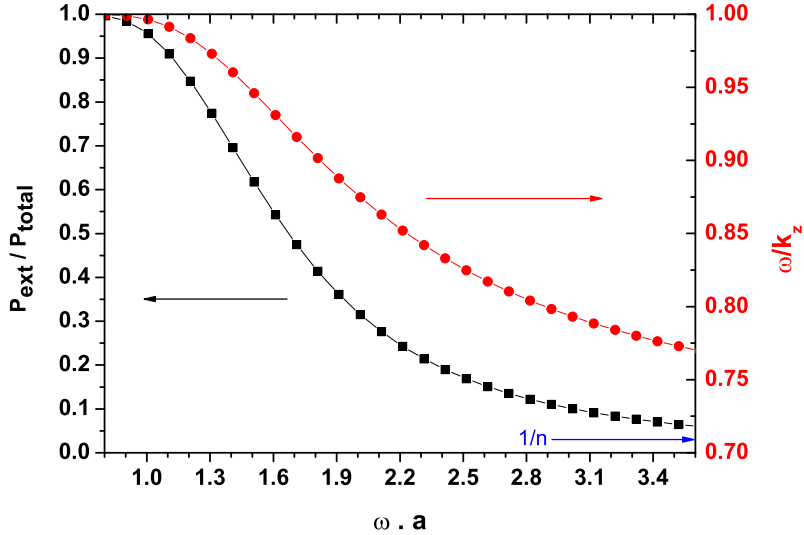}		
	\label{fig:FIG1}
\caption{external fraction of the power (rectangles, left-hand scale) and phase velocity $v_{ph} = \omega/K$ (balls, right-hand scale) for the $HE_{11}$ mode and $\sqrt{\varepsilon}=1.41$.}
\end{figure}

\subsection{The limit of small crossing angle}

Let us assume that the trajectory is at small angle to the fiber axis:
\be 
v_T \ll v_z\,,\quad \Vv=d\Xv/dt\,,\quad\Vv_T=(v_x,v_y)\,.
\ee
In this case one can neglect the contribution of $d\Xv_T\cdot\Ev_T$ in (\ref{ampli}) and write
\begin{equation}
R_{\omega,m} = {iZe\over\omega} \int dz\, E^*_z(z) \ e^{iz[\omega/v_z-K(\omega,m)]} \,.
\label{ampli-Cher}
\end{equation}
Here $\ E_z(z)$ stands for ${E_z}_{\omega,m}(\rv)$, $\rv$ being a slowly varying function of $z$. From (\ref{distrib'}) the photon number, integrated over $K$, is
\begin{equation}
{\cal N}_{m}= {2Z^2\,\alpha} \int \frac{d\omega}{\omega\, P} \int dz' \, E^*_z(z') \int dz'' \, E_z(z'')  
\ e^{i(z'-z'')(\omega/v_z-K)} \,.
\label{distrib-Cher}
\end{equation}
One makes the change of variables $(z'+z'')/2=z$, $\ z'-z''=\Delta z$. Neglecting the dependence of $E_z^*(z')\cdot E_z(z'')$ on $\Delta z$, the exponent can be integated over $\Delta z$, yielding a factor
$2\pi\delta[K(\omega,m)-\omega/v_z]$. Combined with (\ref{disp}), this $\delta$-function selects discrete values of $K$ (one for each $M$ and $\nu$) such that $v_{ph}=v_z$. Introducing the group velocity 
%\be
$v_g=d\omega/dK$
%\ee
of the mode, one can rewrite (\ref{distrib-Cher}) as
\begin{equation}
{\cal N}_{m}= 4\pi Z^2\alpha \, \, {v_z\over P \,\omega} \, {1\over |1- v_z/v_g|} \int dz \, \left|{E_z}_{\omega,m}(\rv)\right|^2 \,
\label{distrib-Cher'}
\end{equation}
(the second fraction before the integral can be rewritten as $d\ln(v_{ph})/d(\ln\omega)$). 
This result corresponds to the "fiber Cherenkov radiation" studied in \cite{Bog-Bol,Zhev-Gleb-Cherenkov}. It applies also to the case of slightly bent trajectory or slightly bent fiber. Equation (\ref{distrib-Cher'}) tells that the photon number increases linearly with the path length over which the particle is close to or inside the fiber. 

The ``fiber Cherenkov radiation'' can be used to measure the velocity of a semi-relativistic particle beam, using the dependence of $v_{ph}$ on $\omega$ shown in Fig.1.

\begin{figure}[htbp] 
	\centering
\includegraphics*[scale=0.5,clip,bb=50 300 550 800]{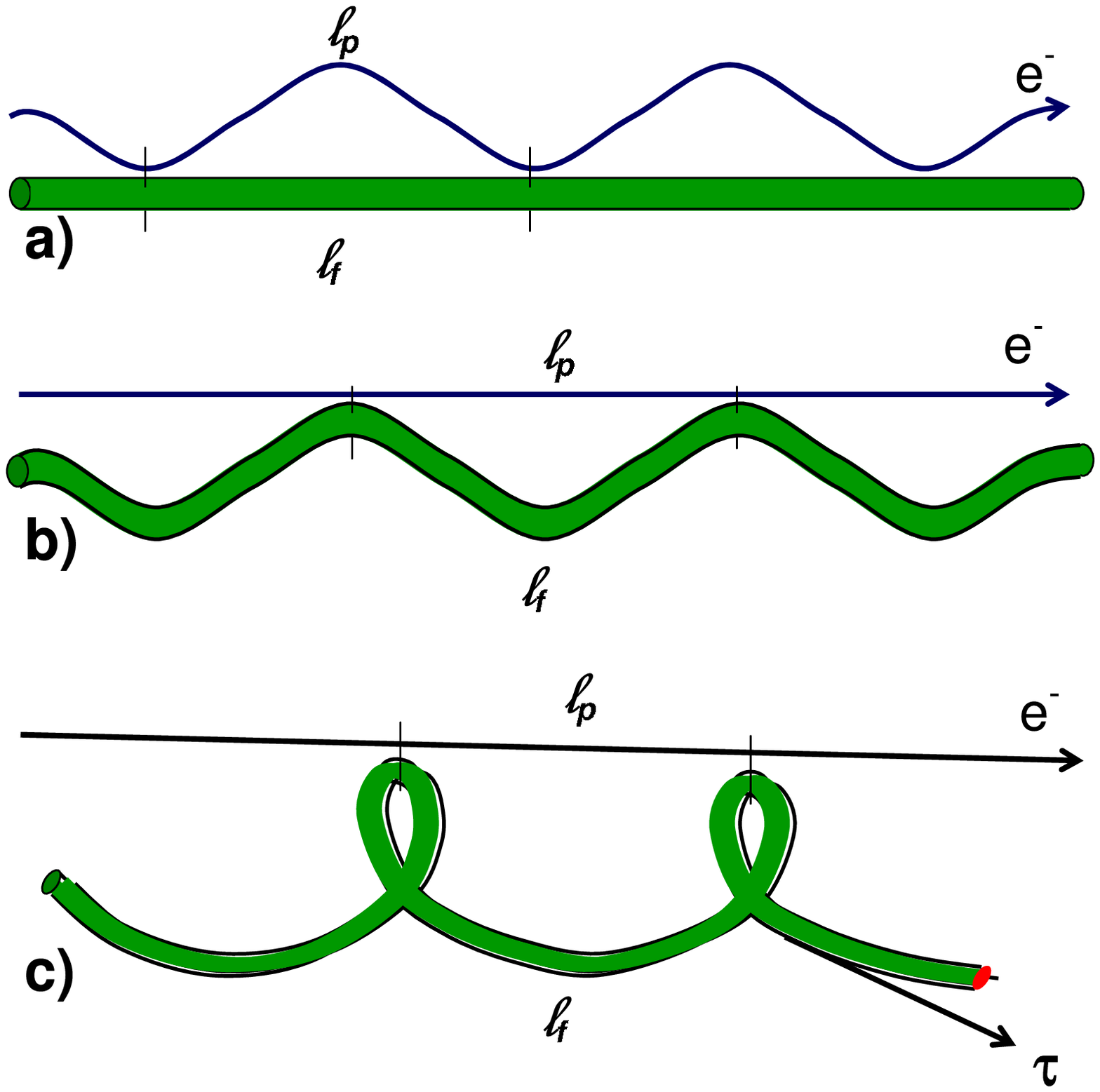} 
	\label{fig:FIG2}
\caption{periodically bent trajectory (a) or bent fiber (b and c). $l_p$ and $l_f$ are the lengths of the curved or straight periods, for the particle and the fiber respectively.}
\end{figure}

\subsection{Interferences with periodically bent trajectory or bent fiber}
Bending the particle trajectory (Fig.2a), the fiber (Fig.2b) or both, one can make several crossing points at which radiation can be produced by polarisation of the medium. Let $l_f$ and $l_p$ be the fiber and trajectory lengths between two crossing points $C_1$ and $C_2$. The particle passes through these points at times $t_1$ and $t_2$. The waves produced at $C_1$ and $C_2$ get a phase difference $\Delta\Phi$ due to the different times of emission and different optical paths. If the fiber is bent in a plane, 
\be
\Delta\Phi=K \, l_f - \omega \, (t_2-t_1) 
%= K \, l_f - \omega \, l_p/v \,
 =\omega \, \left(l_f/v_{ph}-l_p/v\right) \,
\ee
(here $K$ is the wave number along the local fiber axis). If $N$ crossing points, with the same crossing angle and impact parameter, are spaced periodically, the frequency spectrum is 
\begin{equation}\label{N-crossing}
\left({d{\cal N}_{\omega,m}\over d\omega}\right)_{N\, crossing} 
= \left({d{\cal N}_{\omega,m}\over d\omega}\right)_{one\, crossing}
\times\frac{\sin^2(N\Delta\Phi/2)}{\sin^2(\Delta\Phi/2)}\,.
\end{equation}
The interference factor on the right is the usual one, found for instance in X-ray transition radiation from regularly spaced foils. For large $N$ it gathers the photons in quasi-monochromatic lines of frequencies such that 
\be
\Delta\Phi=2k\pi \quad (k\ {\rm integer})\,.
\ee
 If the fiber bending is not planar, for instance helicoidal (Fig.2c), the phase velocities of left- and right-handed polarisations in the fiber are split and the preceding condition becomes
\begin{eqnarray}
	\Delta\Phi = 2k\pi \pm \phi_B
\end{eqnarray}
where $\phi_B$, called {\it Berry phase}, is equal to the solid angle of the cone drawn by the tangent $\tau$ to the fiber \cite{TomitaChiao}.

The observation of interferences may provide a tool for measuring the angular spread of a charged particle beam: the lines are broadened or may disappear if the impact parameters vary from one crossing point to the other.  

\begin{figure}[htbp] 
	\centering
\includegraphics*[angle=-90,scale=0.4,clip,bb=80 30 430 830]{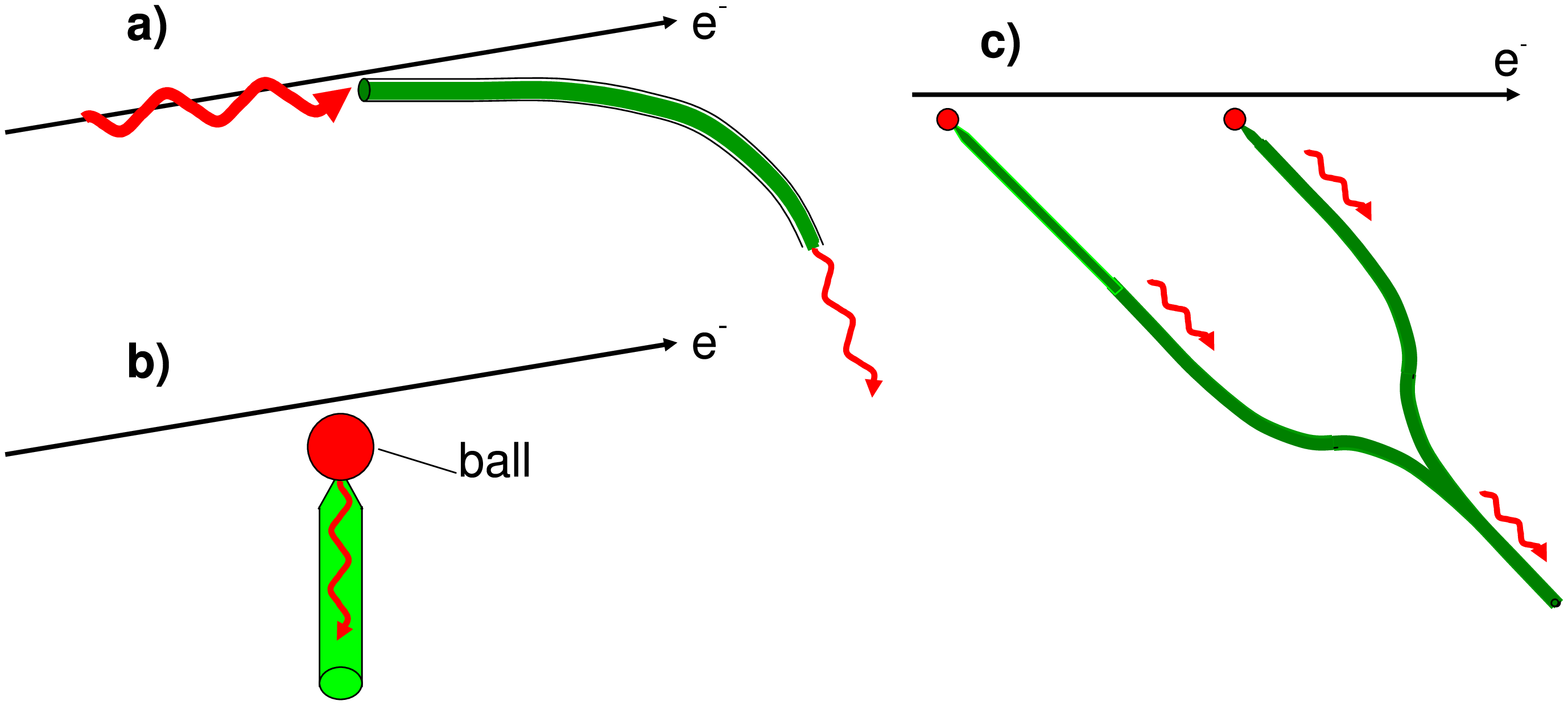}
	\label{fig:FIG3}
\caption{capture of a virtual photon in a non-continuous part of the fiber. a) through the end section; b) through a metallic ball at the fiber end, via a plasmon excitation; c) through two balls. In the latter case, the lights collected by the two balls add coherently at the junction. In the figure the diameter of the metallic ball ($\lesssim\lambdabar\sim10^2$ nm) has been exagerated relative to the fiber diameter  ($\lesssim5\lambdabar$ for a monomode one).}
\end{figure}

\section{Production of light at non-continuous parts of the fiber}\setcounter{equation}{0}

The entrance section of a cut fiber can catch free real photons and convert them into guided photons. Assuming that the photons are nearly parallel to the fiber, the energy spectrum captured by the fiber in the mode $m=\{M,\nu\}$ is approximately given by
\begin{eqnarray}\label{capt-spect}
{dW_m\over d\omega}={1\over4\pi P{(\omega,m)} } \times 
\qquad\qquad\qquad\qquad\qquad\qquad\qquad\qquad\qquad\qquad \cr
\left| \int  d^2\rv \, \left[
 T_B(\rv)\,{\et}^*_{\omega,m}(\rv)\times{\bt}^{in}_\omega(\rv) 
+T_E(\rv)\,{\et}^{in}_\omega(\rv) \times {\bt}_{\omega,m}^*(\rv)
\right]\right|^2 \,.
\end{eqnarray}
where $\{\et^{in},\bt^{in}\}$ is the incoming field and $T_E(\rv)$, $T_B(\rv)$ are the Fresnel refraction coefficients at normal incidence, given by
\be
T_E(\rv)= 2/(1+\sqrt{\varepsilon(\rv)})\,, \quad T_B(\rv)= \sqrt{\varepsilon(\rv)}\,T_E(\rv) \,;
\ee
$T_E(\rv)=T_B(\rv)=1$ outside the fiber. $P(\omega,m)$ is the mode power given by (\ref{flux}). A derivation of (\ref{capt-spect}) is given in Appendix A. This formula can be applied to the capture of virtual photons from the Coulomb field of an ultrarelativistic particle passing near the entrance face (see Fig.3a), since these photons are quasi-real (it does not apply to the case where the particle itself enters the fiber, since the photons at small transverse distance from the particle are too much virtual). 
%The Fourier components of the Coulomb field 
${\et}^{in}_\omega(\rv)$ for the Coulomb field is given in many textbooks in terms of modified Bessel functions, e.g., in Eq.(13.29) of \cite{Jackson} or (19-32) of \cite{Panofsky}. The associated magnetic field is ${\bt}^{in}_\omega(\rv)=\vv\times{\et}^{in}_\omega(\rv)$.

It is also possible to catch virtual photons by a metallic ball, where plasmons are created \cite{Zhev-plasmon,Garcia+Y+A}. Then a fiber sticked to the ball on the extremity (Fig.3b), or on the side as in Fig.4, can evacuate the plasmon in the form of guided light. A rough estimate of the energy stored into plasmons can be obtained when the impact parameter of the particle is large compared to the ball radius $R$ and the time scale $\Delta t\sim b/(\gamma v)$ of the transient field is short compared to the reduced plasmon period $1/\omega=\lambdabar$~($\lambda=2\pi\lambdabar$ is the corresponding photon wavelength in vacuum): the particle field boosts each electron of the ball with a momentum $\qv\simeq2Z\alpha\,\bv/(vb^2)$. It results in a collective dipole excitation of the electron cloud, of energy
\be
W(b)\simeq {4\pi R^3 n_e \over3}\, \left({2Z\alpha\over vb}\right)^2 \, {1\over2m_e}
= {2 Z^2\alpha\over3v^2}\, {\omega_P^2 R^3 \over b^2}
\quad(R\ll b \ll \gamma v/\omega) \,,
\ee
where $\omega_P=(4\pi\alpha n_e/m_e)^{1/2}$ is the plasma frequency of the infinite medium. 
For a spherical ball the dipole plasmon frequency is given by simple formula $\omega=\omega_P/\sqrt{3}$, assuming the Drude formula $\varepsilon^2=1-\omega_P^2/\omega^2$ and neglecting the  retardation effects (case $R\lesssim\lambdabar$). The number of stored quanta is then
\be
{\cal N}(b)={W(b)\over\omega}\simeq {2 Z^2\alpha\over v^2} \cdot { R^3\over\lambdabar b^2} \,.
\ee
Taking $b_{min}=R$ and $b_{max}=\gamma v\lambdabar$,  the cross section for this process is
\be\label{log-bmax}
\sigma=\int_{b_{min}}^{b_{max}} 2\pi\,b\,db \, {\cal N}(b)
\simeq
{4 Z^2\alpha\over  v^2}\cdot { R^3\over\lambdabar} \cdot 
\ln {\gamma v \lambdabar\over R} \,.
\ee
More precise theoretical values of the plasmon frequencies, or experimental ones, are used in \cite{Zhev-plasmon,Garcia+Howie,Garcia+Y+A} in the context of Smith-Purcell radiation. Elliptical balls or bumps are considered in \cite{Zhev-plasmon,Zhevago+Glebov}. Retardation effects and other mutipoles are taken into account in \cite{Garcia+Howie,Garcia+Y+A}. 
%The condition $b_{min}<b_{max}$ implies $R<\gamma v/\omega=\gamma v\lambdabar$. 
A typical order of the cross section, $\sigma\sim10^{-2}\lambdabar^2$ is obtained with $R\sim\lambdabar$, $\ Z=1$, $\ \gamma v\sim1$.  The plasmon wavelength is typically $\lambdabar\sim10^2$nm. Larger cross section can realized  by increasing $R$, but higher multipoles will dominate, unless $\gamma$ is increased simultaneously. Discussions and experimental results about this point are given in \cite{Garcia+Y+A}. 

Once a plasmon is produced, it is not necessarily transmitted to the fiber.  It may be radiated in vacuum or decay by absorption in the metal. Thus the efficiency of the ball scheme depends on the transmission coefficient. We have no information yet about this important parameter. 

\paragraph{Interferences in the ball and cut-fiber schemes.} 

Two cut fibers can be gathered in a single fiber through a junction, making a kind of a pitchfork (see Fig.3c). When a particle passes near the ends of the two branches, the emitted waves add coherently. One can adjust the  branche lengths so that the light signals from the two balls arrive at the junction simultaneously. In Fig.3c this requires that the angle between the branches and the trajectory is equal to the Cherenkov angle corresponding to the light velocity $v_{ph}$. Then, provided that the impact parameters at the two balls are equal, there is a constructive  interference which enhances the total photon number by a factor 2. The observation of this interferences may provide a sensitive test of the angular spread of the beam. 

One may also stick metallic balls at equal spacing $l$ on one side of a  fiber (Fig.4a), thus obtaining constructive interference peaks given by the equation 
\be\label{guidedSP}
(\omega/v\pm K) \, l \equiv(1/v\pm1/v_{ph})\, \omega l=2k\pi \,.
\ee
The signs $+$ and $-$ correspond to lights propagating backward and forward respectively. Equations (\ref{disp}) and (\ref{guidedSP}) fix $\omega$ and $K$. This process is in competition with the Smith-Purcell radiation from the balls, where $\pm1/v_{ph}$ is replaced by $\cos\theta_{rad}$. We can call it "guided Smith-Purcell" radiation. It is advantageous to choose $L$ such that $\omega$ lies on a plasmon resonance of the ball. 
\begin{figure}[htb] % ou [ht]
	\centering
\includegraphics*[scale=0.5,clip,bb=10 160 600 730]{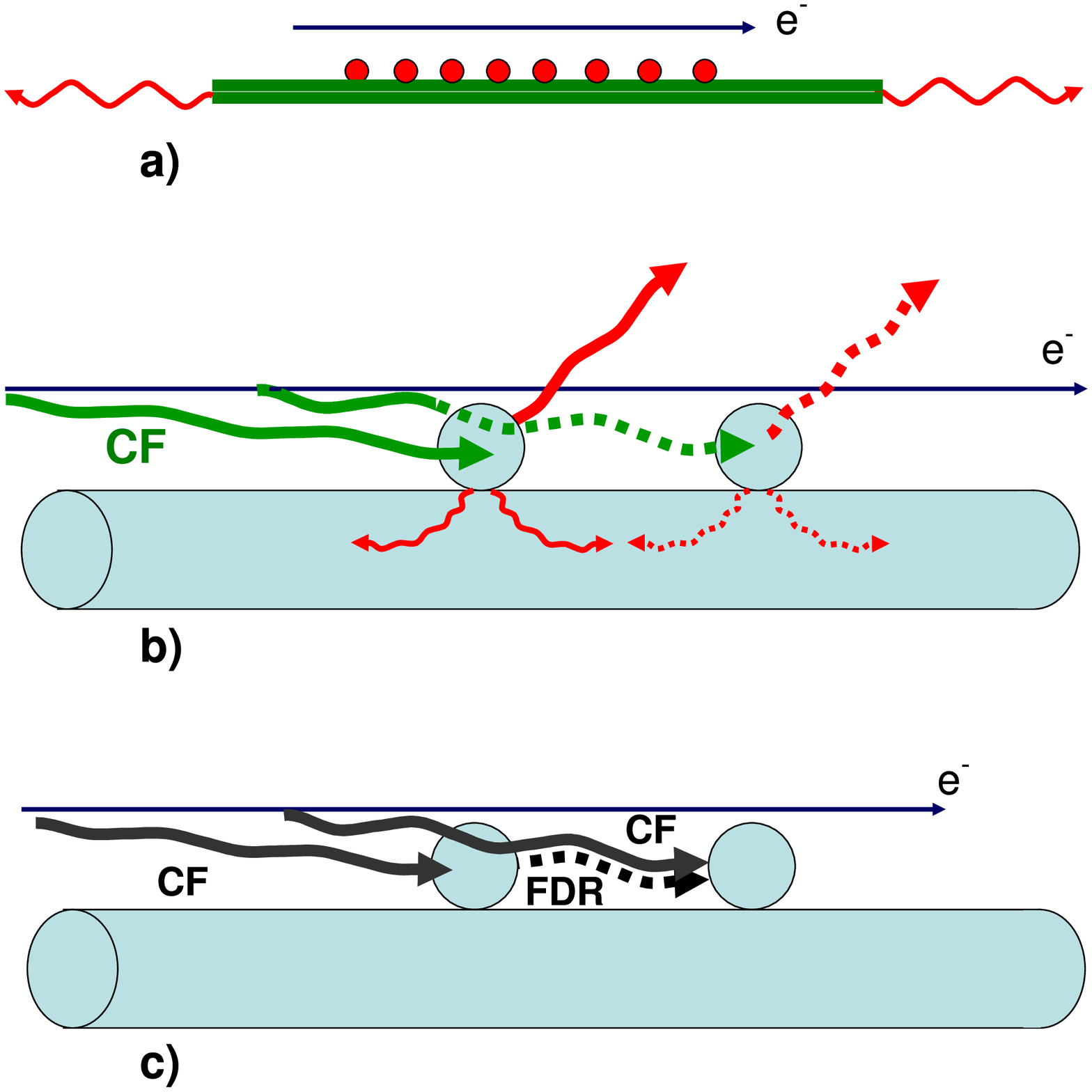}		
	\label{fig:FIG4}
\caption{production of guided light via several metallic balls sticked to a continuous part of the fiber.
a) string of equally spaced balls; b) shadowing of a ball by another one; ``CF'' represents a virtual photon of the Coulomb field of the particle; c) explanation of the shadowing by a destructive interference between the Coulomb field (CF) and the forward diffraction radiation (FDR).}
\end{figure}
\paragraph{Respective advantages of the schemes of sections 2 and 3.}

A real photon (i.e., coming from a distant source) impinging on a continuous part is just scattered, not trapped. Indeed, this photon is in the continuum of the radial quantum number $\nu$, which is conserved due to the translation invariance along the fiber axis. Thus the continuous fiber only acts as a {\it near field} detector. The advantage of this scheme is that it does not receive background from light produced by upstream parts of the accelerator (e.g., synchrotron light from an upstream magnet or diffraction radiation from a collimator). 

In the non-continuous fiber scheme (cut fiber or sticked metallic ball), real photons can be captured. The heterogeneity acts both as a {\it near field} and {\it far field} detector. One must therefore take into account the background from upstream radiation. 
On the other hand, the efficiency of this scheme increases when the particle becomes ultrarelativistic,  due to the growth of the quasi-real photon cloud; it gives the logarithmic factor in (\ref{log-bmax}).  
No such logarithmic rise is expected in the continuous fiber scheme. 

As regards the resolution power in beam size measurement, in the ultrarelativistic case one has to take into account the transverse expansion of the virtual photon cloud, whose mean transverse size grows like $\gamma\lambdabar/\ln\gamma$ (if upstream magnets are far enough). This may lead to a somewhat poorer resolution, but experience with optical transition radiation monitors shows that this effect is not drastic. As for the method where light is produced in the continuous part of the fiber, it is not sensitive to the cloud expansion, since the latter concerns only quasi-real photons.

\section{Shadowing}\setcounter{equation}{0}

It is tempting to write the guided Smith-Purcell photon yield from $N$ balls in a form similar to (\ref{N-crossing}). However, like with real photons, the flux of virtual photons after the first ball is reduced. In other words, each ball makes a shadow for the next balls (see Fig.4b). Let the amplitude of the wave produced by the $n^{th}$ ball be $A_n=A_1 \cdot S_n$ where $A_1$ is the amplitude for one ball only, $S_1=1$ and $|S_n|<1$ for $n>1$. At large $N$, the shadowing factor $\ S_N$ tends toward an asymptotic value $S$ and we can write
\begin{equation}\label{shadowfactor}
\left({d{\cal N}_{\omega,m}\over d\omega}\right)_{N\, ball} 
\simeq \left({d{\cal N}_{\omega,m}\over d\omega}\right)_{one\, ball}
\times\frac{\sin^2(N\Delta\Phi/2)}{\sin^2(\Delta\Phi/2)} \times |S|^2\,.
\label{N-crossing+shadow}
\end{equation}
Shadowing between two balls can also be viewed as a destructive interference between the unperturbed particle field and the forward diffraction radiation from the first ball as schematised in Fig.4c. It belongs to the rescattering effects studied by Garc\'ia and co-authors \cite{Garcia-99+00}. 
The shadow region behind a ball ceases when the diffraction radiation and the particle field get out of phase. Thus shadowing acts in the so-called \emph{formation zone} of lenght $l_f\sim\lambdabar/(1-v)$. This is the necessary distance to restore the virtual photon cloud. Effects of incomplete virtual photon cloud have already been pointed out by Feinberg \cite{Feinberg}, Fomin and Shul'ga \cite{Fomin}. 
Shadowing is most important with ultrarelativistic particles, the formation zone $l_f\sim\gamma^2\lambdabar$ becoming much longer than the ball spacing. 

\paragraph{A possible universal bound for the energy loss by the particle passing near a radiator.}
{}\smallskip

The necessity of ``repairing'' the virtual photon cloud after each ball probably leads to a bound on the energy loss per unit lenght in the device shown in Fig.4a. This should be the case for all types of periodic \emph{radiator} (e.g., for transition radiation, Smith-Purcell radiation and parametric X-ray%
\footnote{In \emph{diffracted transition radiation}, the shadowing corresponds the ``dynamical'' effect discussed after Eq.(28) of \cite{XA-Rul}.
%properly taken into account, in the dynamical theory, by the density effect and Bragg attenuation (
}) % 
based on the polarisation of the medium, in which the particle is moving in a linear uniform motion. Here we consider the case where the particle does not touch the medium but stays at a minimum distance $b$ from it. 
A first guess of the maximum energy loss is based on the following classical argument:
A cylinder of radius $b$ around the trajectory contains no mater. On the cylinder surface, the radial component of the Poynting vector (Appendix B) of the particle field takes the value 
\be\label{Pr}
{\cal P}_r(t,z)=-{Z^2\alpha\over4\pi}\,\gamma^2vb\ (z-vt)\, \left[b^2+\gamma^2\,(z-vt)^2\right]^{-3}\,.
\ee
Thus, in front of the particle ($z>vt$), electromagnetic energy is flowing inwards the cylinder and behind the particle ($z<vt$), it is flowing outwards. Let us make the assumption that the maximum power that the radiator can extract from the particle is the power $P_{behind}$ flowing outwards behind the particle. It gives the following bound for the linear energy loss: 
\be\label{Bound1}
{-dE\over dz} \le {P_{behind}\over v} = {1\over8} \cdot {Z^2\alpha\over b^2} \,.
\ee
For the Smith-Purcell case we may replace the cylinder by a plane at distance $b$ from the trajectory. A smaller bound is then obtained: ${-dE/dz}\le Z^2\alpha/(32b^2)$. 

A weak point of the above assumption is that the Poynting vector of Eq.(\ref{Pr}) does not include the field scattered by the radiator and the interference term between this field and the particle field. Nevertheless a similar bound can be derived from the formation zone effect. Let us consider an ultrarelativistic particle passing through a series of holes of radius $b$ in material foils. The first foil absorbs the virtual photons of impact parameters larger than $b$, or reflects them as backward diffraction radiation. These photons carry an energy 
\be\label{W1}
W_1= {3\pi\over16} \cdot {\gamma Z^2\alpha\over b} \,,
\ee
which is lost for the particle. If, instead of pierced foils, the radiator consists in half-plane foils whose edges are at distace $b$ from the beam, $W_1$ is given by the same formula without the factor $\pi$. The calculation of $W_1$, that we define as the flux of the Poynting vector on the foil, integrated over time, is given in Appendix B. In addition, forward diffraction radiation is emitted and also takes out the energy $W_1$ from the particle. Without shadowing, the energy loss per unit length in the radiator would be $-dE/dz=2W_1/l$ for a foil spacing $l$. The virtual photons have a flat energy spectrum up to a cut-off energy $\omega_{max}\sim\gamma/b$, therefore their formation lengths are at least $\gamma^2/\omega_{max}\sim\gamma b$. Due to shadowing, reducing the foils distance to less than $\gamma b$ does not increases the linear energy loss. We obtain thus the upper bound
\be\label{Bound2}
{-dE\over dz} \lesssim {3\pi\over8} \cdot {Z^2\alpha\over b^2} \,.
%{\pi Z^2\alpha\over32b^2} \,
\ee
Owing to the non precise definition of the formation length ($\sim \gamma^2\lambdabar$ or $\gamma^2\lambda$), this result is not incompatible with (\ref{Bound1}). This supports the hypothesis of the existence of an upper bound for $-dE/dz$.

The conjectured bound concerns the total energy loss (radiation + energy deposit in the material). It 
should apply as well to the Smith-Purcell radiation, the ``Cherenkov at a distance'' effect and to a particle passing along a rough non-periodic surface. It would be of great interest to prove (or disprove) its existence rigorously and, if it exists, to refine the numerical coefficient. This coefficient is expected to be smaller, probably by a factor $3-4$, for a radiator located behind a plane (usual Smith-Purcell case) than around a cylinder. 

\section{Outlook}\setcounter{equation}{0}

In this paper we have described several ways by which charged particles produce, via polarisation of the medium, coherent light in narrow optical fibers: crossing or passing near a continuous part of the fiber, passing near an end or near a metallic ball sticked to the fiber. We have discussed some of the interference effects occuring when the particle produces light at several points of the fiber.  
For a particle moving in uniform motion along a row of balls, we have pointed out a possible reduction of the photon yield by the shadow effect. It led us to conjecture an upper bound for the linear energy loss $-dE/dz$, depending only on the impact parameter and which may apply to other sources of radiation as well, for instance to Smith-Purcell radiation.  

We have given two examples of application to beam diagnostics: 

%\noindent 
- measurement of the particle velocity, using the ``fiber Cherenkov effect'' with straight parallel beam and fiber,

- measurement of the angular spread, with the interference of the radiations collected from two or more aligned metallic balls. 

\noindent 
Much work remains to be done to estimate the photon yield which can be obtained with these different mechanisms: find the ball-to-fiber transmission coefficients, study the photon polarisation, choose the most convenient wavelength (infra-red, visible or ultraviolet), fiber diameter, ball diameter, etc. 
The intensity may be too weak for single particle detection (contrary to the thick fiber case of Refs.\cite{DIRC,Gorodetsky}), but enough for beam diagnostics. 

From the technical point of view, the flexibility of a fiber can be taken as an advantage, mut may rise mechanical problems. The fiber has to be narrow if one wants to make use of the interference effects or the monochromatic  fiber Cherenkov radiation, otherwise too many transverse modes are excited. Besides, a narrow fiber has less effects on the beam emittance.

The existence of an upper bound for $-dE/dz$ is also a theoretical question. 

\medskip
We thank Mr. Alexei Tishenko for comments and Prof. Gennadi Naumenko for pointing us Refs.\cite{Feinberg,Fomin} and showing us preliminary experimental results on the shadowing effect.

\section{Appendix A. Derivation of Eq.(\ref{capt-spect})}\setcounter{equation}{0}

\paragraph{1. Orthogonality relations between modes.}
We define the \emph{power scalar product} of two fields $\Fv=\{\Ev,\Bv\}$ and $\Fv'=\{\Ev',\Bv'\}$ at a given $z$ as
\be\label{power-scalar}
(\Fv|\Fv') = (\Fv'|\Fv)^* = {1\over2}\int d^2\rv \ \left[\, 
\et^*(\rv)\times\bt'(\rv) + \et'(\rv)\times\bt^*(\rv) \,\right]\,.
\ee
For proper transverse modes, $\Gv_{\omega,m}=\{\Ev_{\omega,m},\Bv_{\omega,m}\}$,
one has the orthogonality relation at given $\omega$: 
\be\label{ortho}
(\Gv_{\omega,m}|\Gv_{\omega,n})=P(\omega,m) \ \delta_{mn}\,,
\ee
where $P(\omega,m)$ is the power of mode $m$ calculated with (\ref{flux}). To prove (\ref{ortho}) one may consider the \emph{longitudinal momentum operator} $\K$, acting in the given $\omega$ subspace and defined by
\be
\Fv=\pmatrix{\et \cr\el \cr\bt  \cr\bl} \longrightarrow
\K\,\Fv=\pmatrix{
-\omega\,\zu\times\bt-i\nabla_T\el \cr
{i\over\varepsilon(\rv)}\nabla_T(\varepsilon(\rv)\et) \cr
\omega\varepsilon(\rv)\,\zu\times\et-i\nabla_T\bl \cr
i\nabla_T\cdot\bt
}\,.
\ee
Its eigenvalues are the longitudinal momenta $K$ (the eigenstate equation $\K\,\Fv=K\,\Fv$ is equivalent to the Maxwell equations). $\K$ is hermitian with respect to the power scalar product, that is to say
\be
\left(\Fv|\K|\Fv'\right)=\left(\Fv'|\K|\Fv\right)^* \,.
\ee
It results that at fixed $\omega$ two different modes (which have different values of $K$) are orthogonal with respect to the power scalar product. This justifies (\ref{ortho}). 

\paragraph{2. Decomposition of a light signal into modes.}

Introducing the time Fourier transform of $\Fv(\rv,0,t)$, 
\be
\Fv_\omega(\rv) =
\int^{\infty}_{-\infty} dt\ e^{i\omega\,t}\ \Fv(\rv,0,t)  \,,
\ee
we can get from Eq.(\ref{decompose-mode}) 
\be\label{F=sommode}
\Fv_\omega(\rv) = \sum_{m} 
c{(\omega,m)}  \  \Gv_{\omega,m}(\rv)  \,,
\ee
where, from orthogonality, the coefficient $c{(\omega,m)}$ is given by
\be
c{(\omega,m)} ={ (\Gv_{\omega,m}|\Fv_\omega) \over  P(\omega,m) }
\ee
\be
= {1\over2P{(\omega,m)} }\int  d^2\rv \ \left\{
  \Ev_{\omega,m}^*(\rv)\times\Bv_{\omega}(\rv)
+\Ev_{\omega}(\rv) \times \Bv_{\omega,m}^*(\rv)
\right\}_z \,.
\ee
The energy of the field $\Fv(\rv,t)$ flowing through the plane $z=0$ from $t=-\infty$ to  $t=+\infty$ is
\be
W=\int^{\infty}_{-\infty} dt \ (\Fv(\rv,t)|\Fv(\rv,t)) \,
\ee
(in this expression, $\rv$ is a dummy variable, integrated over in (\ref{power-scalar})). 
After some algebra, and using the orthogonality relation (\ref{ortho}) one gets
\be
W=\frac{1}{\pi}\sum_{m} \int^{\infty}_{0} d\omega \
|c{(\omega,m)}|^2  \ P(\omega,m) \,.
\ee
The energy spectrum captured from the field $\Fv=\{\Ev,\Bv\}$ in mode $m$ is then
\be\label{partageSpectre}
{dW_m\over d\omega}= {1\over4\pi P{(\omega,m)} }
\left| \int  d^2\rv \ \left\{
 \Ev_{\omega,m}^*(\rv)\times\Bv_{\omega}(\rv)
+\Ev_{\omega}(\rv) \times \Bv_{\omega,m}^*(\rv)
\right\}\right|^2 \,.
\ee
Equation (\ref{capt-spect}) differs from the preceding one just by the Fresnel coefficients $T_E(\rv)$ and $T_B(\rv)$ which take into account the refraction of the incoming field at the entrance face. 

\section{Appendix B. Derivation of Eqs.(\ref{Bound1}-\ref{W1})}\setcounter{equation}{0}

The electric field of the particle and the Poynting vector are given in cylindrical coordinates by 
\be\label{}%(\ref{})
%\Ev=
\pmatrix{E_r\cr E_\phi\cr E_z}={\gamma Ze\over4\pi(r^2+\gamma^2\zeta^2)^{3/2}}\pmatrix{r\cr0\cr\zeta}\,,\quad 
{\cal P}=\pmatrix{-vE_rE_z\cr0\cr vE_r^2}
\,,\ee
with $\zeta\equiv z-vt$ (the magnetic field is $\Bv=\vv\times\Ev$).
Integration over ${\cal P}_r$ on the cylinder $r=b$ and in the $z$ range $[-\infty,vt]$ yields the result (\ref{Bound1}). A 4 times smaller quantity is obtained if one integrates over the half-plane $x=0$, $y=b$, $z\le vt$. 

Integration over ${\cal P}_z$ on the plane $z=0$ minus the hole $r<b$ and in the time range $[-\infty,+\infty]$ yields the result (\ref{W1}). A $\pi$ times smaller quantity is obtained if one integrates over the half-plane $z=0$, $y\ge b$. 

\section{FIGURE CAPTIONS}

Fig.1: external fraction of the power (rectangles, left-hand scale) and phase velocity $v_{ph} = \omega/K$ (balls, right-hand scale) for the $HE_{11}$ mode, for $\sqrt{\varepsilon}=1.41$.

Fig.2: periodically bent trajectory (a) or bent fiber (b and c). $l_p$ and $l_f$ are the lengths of the curved or straight periods, for the particle and the fiber respectively.  

Fig.3: capture of a virtual photon in a non-continuous part of the fiber. a) through the end section; b) through a metallic ball at the fiber end, via a plasmon excitation; c) through two balls. In the latter case, the lights collected by the two balls add coherently at the junction. In the figure the diameter of the metallic ball ($\lesssim\lambdabar\sim10^2$ nm) has been exagerated relative to the fiber diameter  ($\lesssim5\lambdabar$ for a monomode one).

Fig.4: production of guided light via several metallic balls sticked to a continuous part of the fiber.
a) string of equally spaced balls; b) shadowing of a ball by another one; ``CF'' represents a virtual photon of the Coulomb field of the particle; c) explanation of the shadowing by a destructive interference between the Coulomb field (CF) and the forward diffraction radiation (FDR). 

\end{document}